\documentstyle[preprint,aps,epsfig]{revtex}
\begin{document}
\newcommand{\mex}[1]{$ \langle #1 \rangle $ }

\title{
New features of dislocation structures arising
from  lattice discreteness.
 }
\author{
Oleg N. Mryasov$^{1,2}$, Yu. N. Gornostyrev$^{2,3}$ and A.J. Freeman$^{2}$
}
\address{
\(^1\)  Department of Materials Science, University of California,
 Berkeley, CA 94720-1760. \\
\(^2\)  Department
of Physics and Astronomy, Northwestern University Evanston, IL
60208-3112. \\
\(^3\) Institute of Metal Physics, Ekaterinburg, Russia.
}
\maketitle

\begin{abstract}
New aspects of a relation between lattice and dislocation structures are
examined
within a physically transparent theoretical scheme.
Predicted features originating from the lattice  discreteness  include:
(i) multiple core  dislocation structures and
(ii) their dependence on the position of the dislocation axis.
These effects, which in principle can be observed directly and may also
manifest themselves in dislocation motion or/and transformation (cross-slip)
characteristics, are very general and  present in any crystal
in which they may be more or less pronounced depending on the material.

\end{abstract}


It is widely accepted that such defects as dislocations
significantly influence a number of properties in real materials.
Thus, understanding the  relationship between  lattice and dislocation
structures
is one of the  fundamental problems of  materials physics.
Despite  recent developments of powerful atomistic simulation techniques,
up to now
our understanding of
the relation between lattice
and dislocation structures
is based on
results obtained within the framework of the
Peierls-Nabarro (PN) model\cite{Peierls,HirthLote}.
This model has provided both  language for interpretation
experimental/theoretical results and simple relations between
dislocation properties and lattice discreteness characteristics
(periodicity, symmetry, etc.)
\cite{Peierls,HirthLote,nabb_50y,duePRL}.
This remarkable breakthrough in understanding how dislocation properties
are related with lattice characteristics
 became possible due to two features  of this model:
(i) its high tractability and
(ii) its combined different length scale descriptions.

The combined "continuum/atomistic" descriptions in the PN model
follow clearly from the structure of the energy functionals \cite{schoeck:94}, $E_{tot}$, of the
dislocation displacement distribution $\vec{u}(x)$
(here $x$ is a distance from the dislocation axis in the slip plane)
\begin{eqnarray}
E_{tot}(\vec{u}(x))= E_{el} (\vec{u}(x))  + E_{mis} (\vec{u}(x))
\label{funct}
\end{eqnarray}
with  a linear elastic ($E_{el}$ ) and a non-linear atomistic
 misfit energy term ($E_{mis}$)
\begin{eqnarray}
E _{mis}= h \sum \limits _n \Phi (\vec u (nh-l)) ,
\label{eq:misf}
\end{eqnarray}
where $\Phi (\vec u (x))$ is  a  periodic energy profile
which is often approximated by
the so-called generalized
stacking fault energy (GSF) or $\gamma$-surface \cite{vitekGSF}.
Indeed, the $E_{mis}$ term represents the most apparent and important
lattice properties
- discreteness/periodicity/symmetry - and allows one to investigate
within the PN model the relation between lattice and dislocation properties.
$E_{mis}$
can be expanded in a Fourier series  as
\begin{eqnarray}
E _{mis} = E_{mis}^0 + h  \sum \limits _{s=1}^{s=\infty} J_s \cos \frac
{2\pi sl}{h}
\label{eq:misf_ft}
\end{eqnarray}
where $E_{mis}^0$ $= \int \limits _{-\infty}^{\infty}\Phi (\vec u (x))dx$
is independent of the position of the dislocation axis  $l$
and  terms which are oscillatory  with  $h$  being a repeat distance normal
to the dislocation line
and
$n$  an integer number that counts atomic rows in the same direction.

Now, a minimization of ${E_{tot}(\vec{u}(x))}$ allows  one
to find the equilibrium dislocation structure,   $\vec{u}(x)$.
In order to perform  this minimization in analytic form,
a critical approximation $E _{mis} = E_{mis}^{0}$ has been made
\cite{Peierls}.
This  "continuum" approximation in representing the misfit energy - which
is supposed to describe  lattice discreteness - results in an obvious
inconsistency
which has been a subject of debate for years,
mainly  in the context of  Peierls stress determinations
\cite{nabb_50y,bulatov_l,ohsawa1}.
Thus, an interesting and fundamental  issue  arises -  if
consistently represented in the atomistic interaction energy, how will
lattice discreteness   be manifested
in  the structure of dislocations ?

Several recent attempts to overcome this inconsistency
resulted in
purely numerical procedures \cite{bulatov_l,ohsawa1}. Thus, one of the most
advantageous  features of the PN model -
high tractability and transparency - has been sacrificed.
These authors also focused on the Peierls stress determination and
demonstrated that
indeed the discrete representation of the misfit energy
brings theoretical estimates  much closer to experimental results.

In this Letter,   using
a physically transparent solution of the PN model with
a consistent discrete representation for
the misfit energy, we
examine how lattice discreteness
may influence dislocation structure.
This allows us to predict
new generic features
of the dislocation structure that are independent of the PN model assumptions,
and driven by  lattice discreteness
such as multiple core structures
and their dependence on the position of the dislocation axis.

To determine dislocation structure,
we perform a minimization  of the  total energy  functional,
Eq. \ref{funct}, with  a discrete representation of the
misfit energy, Eq.\ref{eq:misf}, using trial functions, $\vec{u}(x)$,  defined
from the Laurent expansion \cite{Lejcek,prb2}  of their derivatives
$\rho _{\beta}(x)$
\begin{eqnarray}
\rho _{\beta}(x)=\frac{du_{\beta}(x)}{dx} = Re \sum_{k=1}^N
\sum_{n=1}^{p_k} \frac{A_{nk}^{\beta}}
{(x-z_k^{\beta})^n} ,
\label{RhoEqn}
\end{eqnarray}
where
N is the maximal number, $p_{k}$ is the maximal
order of the poles $z_{k}^{\beta}$ and
$A_{nk}^{\beta}$ are expansion coefficients.
It is important to note that,  by definition, these trial functions
provide a minimum of  $E_{tot}$
for an arbitrary $\Phi(u)$ potential (not only sinusoidal, 
as trial functions used in \cite{schoeck:94} to parameterize
total energy functional in the convenient form) 
in case of the  "continuum" approximation,
$E _{mis} = E_{mis}^{0}$ \cite{Lejcek,prb2}.
This choice of  trial functions  not only provides good accuracy and
stability of the  minimization procedure \cite{prb2} but also allows
one to express $E_{tot}$ through
parameters describing the dislocation structure.
Indeed, the poles $z_k^{\beta}=x_k^{\beta}+i\omega_k^{\beta}$
have a clear meaning:$x_k^{\beta}$ gives the position and
$\omega_k^{\beta}$ gives  the  width of the
partials for the  screw ($\beta$=1) and edge ($\beta$=2) components of the
displacement in the partial cores.  For example,  for the ordinary
dislocations
dissociated into two Shockley partials,
$x_k^{\beta} = l \pm d^{\beta}/2$, where d is the  partials separation and  $l$
gives the position of the whole ordinary dislocation center.
In the general case with these trial functions, 
 $E_{tot}$
can be presented
as a  numerical function of  geometrical  parameters,
in particular,  for an ordinary dislocation  
as  a function of the set  of parameters ($\{{\bf g
}\}=\{d, \omega, l\}$) describing the dislocation's structure ($d, \omega$)
and  its position in  the lattice ($l$).

As examples, we consider ordinary dislocations  for  fcc metal, Ir,
and an ordered alloy,
CuAu, with L1$_{0}$ structure. In these materials, this type of  dislocation
normally splits into two Shockley partials \cite{HirthLote,prb2}
and represents a very typical example of dislocation structures.
To illustrate  graphically  minimization of $E_{tot}$,
let us introduce elastic ($F_{el}(d)$) and misfit ($F_{mis}(d)$)
generalized forces which are defined
as, \hspace{0.2cm}
$F_{el}(d)= - \partial E_{el}(d)/\partial d$
(the sign is chosen for convenience) and
$F_{mis}(d)= \partial E_{mis}(d)/\partial d$
(here and  further we drop the $\beta$ index  since $d^{1} \approx d^{2}$).
In this definition, for a given $d$
other  geometrical parameters from
the complete set $\{\bf g\}$ are taken
to be such that they minimize $E_{tot}$.
Obviously, in this case the intersection of  $F_{el}(d)$ and $F_{mis}(d)$
gives a partial
separation  $d$ which corresponds to the minimum of Eq.\ref{funct},
provided that
the second derivatives are positive.

The generalized forces calculated according to this definition
using ab-initio $\gamma$-surfaces (see \cite{prb2} for
details)
in the case of the screw orientation
of the unit dislocation for Ir and CuAu are presented in
Fig.\ref{fig:forces} (a,c).
For comparison, we also  determine
generalized forces for dislocations
with simple model  density distribution  displacements composed of two
delta functions,
$\rho (x) = b_{1} \delta(x + d/2)  +  b_{2}\delta(x - d/2)$
(see Fig. \ref{fig:forces} (b, d)).
In this case, we have a step function shaped  dislocation for which
$E_{tot}$
has a very simple   form,
\begin{eqnarray}
E_{tot}=H \cdot ln(\frac{1}{d}) + \gamma_{isf} \cdot d
\label{sing_funct}
\end{eqnarray}
and corresponding  generalized forces
$F_{el}^{step}= H /d $ and  $F_{mis}^{step} = \gamma_{isf} $, where
$ \gamma_{isf} $ is the intrinsic stacking fault energy and H is
a so-called prelogarithmic factor (see for example \cite{schoeck:94}).
Interestingly enough,  for this model type of dislocation
the  well-known simple relation  between the equilibrium partials
separation and
stacking fault energy, $d=H/\gamma_{isf}$ \cite{HirthLote}, can be easily
recovered from
the  functional dependence in Eq. \ref{sing_funct}.
As can be seen in Fig.\ref{funct},
in the limit of large  separation distances ($d>>\omega$),
the PN model generalized forces  defined within the "continuum" approximation
approach
those for the step function shaped dislocation.

We now focus on how the oscillatory part of the misfit energy, usually
neglected in the
PN model analysis, affects dislocation structure.
Remarkably enough, there are not only oscillations with
$l$ which can be expected  from  the  Fourier expansion of the misfit energy,
Eq. \ref{eq:misf_ft},  but also  oscillations with
partials separation $d$ for fixed $l$.
As can be seen in Fig. \ref{fig:forces},
for Ir, despite the rather small amplitude of these oscillations,
the effect of the misfit energy discrete representation  is quite visible since
the intersection of the generalized forces happens to be  in the area
where $d$ dependence of the F$_{el}$ is rather weak and so 
solutions are affected most by the F$_{mis}$ oscillations.
Finally, for CuAu the effect is dramatic since the amplitude
of the oscillations is very large.
It is significant that the discrete representation of the misfit energy not
only
changes quantitatively parameters of the dislocation structure (as in the
case of Ir)
but may also result in qualitatively new effects (as in the  case of CuAu).
As is evident from Fig.\ref{fig:forces}, one such general effect, which is
independent of the PN model approximations,
is the appearance of
multiple stable dislocation core configurations.
Indeed, conclusions we draw in this study about the possibility of multiple
core
configurations are based on generic features of the $F_{el}(d)$ and
$F_{mis}(d)$
dependencies (see Fig.\ref{fig:forces}) which follow from the general physics
of the linear (elastic part) and non-linear (misfit part) lattice response
\cite{note_pn_limit}.

Moreover, within the proposed scheme it is possible to
derive\cite{note_rig_derr}
the following convenient and physically transparent form for the energy
functional, Eq.\ref{funct},
\begin{eqnarray}
E_{tot}= E_{tot}^{0}(d,\omega) +
A(d,\omega) \cos \frac {2\pi l}{h} \cos \frac {\pi d}{h}
\label{tot_nform}
\end{eqnarray}
Here the $l$ independent first term $E_{tot}^{0} = E_{el} + E_{mis}^0 $
is the energy in the ``continuum'' approximation and the second term has an
explicit
oscillatory dependence both on $l$ and $d$.

This form reveals that 
the appearence of  
the above features associated with lattice discreteness
are dependent on the relative contribution of the energies represented by the $E_{tot}^{0}$
and the oscillatory terms.
In turn, the character of $E_{tot}^{0}(d)$ dependence and corresponding 
generalized forces  
is driven by the competition of the partials attraction described by  $E_{mis}^0$ 
( term  
which is dependent upon 
the $\gamma$-surface energetic characteristics and  
for large $d$, it can be well approximated by Eq. \ref {sing_funct})
and the
elastic repulsion ($E_{el}$  which is
dependent on the elastic constants and in the limit $d>>\omega$, has
a simple dependence, see Fig.\ref{fig:forces}).
The influence  of the oscillatory term is predetermined by its amplitude
$A(d,\omega)$ which according to our analysis
is strongly dependent on 
characteristics of the $\gamma$-surface \cite{note_amp}.

It is important that Eq. \ref{tot_nform} describes dislocation energetics
for a wide range of $d$ and $\omega$  \cite{note_schoeck_func}
and correspondingly a complex   interdependence of all geometrical
parameters ($d$, $\omega$ and $l$).
Features of dislocation structure originating in this interdependence of
geometrical parameters
and their impact on dislocation energetics can be  seen from the calculated
$d$ and $l$ dependencies of the $E_{tot}$  in Fig.\ref{fig:compl}.
Indeed, proof that there are can be more than one stable dislocation core
configuration (CuAu)
can be seen in Fig.\ref{fig:compl}(a). Next, the
dependence of the partials separation  on the position of the dislocation
axis in the lattice  is clearly seen in Fig.\ref{fig:compl}(b).
A comparison of the $d(l)$ dependencies determined within the
``continuum'' approximation and with a discrete representation for $E_{mis}$
makes it evident that lattice discreteness is the origin
of the variation in equilibrium dislocation structure depending on
the position of the dislocation axis  (core ``relaxation'').
This variation may result in
changes in the number of stable core configurations
and abrupt transitions between them
(as for CuAu, see Fig.\ref{fig:compl}(a,b)).

These predicted features may have a profound impact
on dislocation energetics.
As can be seen in Fig.\ref{fig:compl}(c),  the dependence
of the dislocation structure on the position of the dislocation axis
may not only lower significantly the Peierls barrier (as also have been
found in \cite{bulatov_l,ohsawa1})
but may even modify the shape of the  Peierls potential.
It is remarkable that core ``relaxation''  along with
the existence of the multiple core configurations (the case of  CuAu)
adds a new feature to the Peierls potential - an additional minimum -
which according to the model  analysis
may result in characteristic changes of the
temperature dependence of the yield stress\cite{takeuchi_doub}.
As can be seen clearly in Fig.\ref{fig:compl},  the  unusual
shape of the Peierls potential  in CuAu
 originates in the abrupt transitions between
two stable core configurations, ``1'' and ``2''.

We find that  in addition to the known dislocation
structure features, 
lattice discreteness is
the origin of (i) multiple core configurations and (ii) their dependence
on the position of the dislocation axis.
Combination of these effects
may result in rather complex variations of the dislocation structure
over the crystal
including
changes in the number of stable core configurations and transitions between them.
As a result, one may have to consider a distribution of the core
configurations
in a crystal under ambient conditions rather than one characteristic
core structure which determines dislocation motion or cross-slip
properties.

As follows from our analysis, predicted  features of dislocation structure
originating
in lattice discreteness  are always present in crystals, but as we demonstrate
with Ir and CuAu as examples, they appear more or less  pronounced
depending on characteristics of the given material.
These fundamental characteristics can be identified within the proposed
theoretical analysis primarily due to  its tractability and physical
transparency \cite{note_amp}.
While these features, namely multiple core configurations,
can be directly verified, in principle,  in high resolution electron
microscopy experiments, they
may also reveal themselves  indirectly in low temperature internal friction
experiments \cite{intf:exp} and in mechanical properties which depend
on elementary processes
that are sensitive to the dislocation structure. Among such processes, we
would emphasize
cross-slip,  where dislocation core structure and its changes under local
stress may play an important role.

Work at Northwestern University was supported by the Air Force Office of
Scientific Research
(Grant No. F49620-95-1-0189)
and at UCB by the Office of Basic Energy Science, Division of
Materials Science, of the U.S. Department of Energy under contract No.
DE-AC04-94AL85000.


\begin{thebibliography}{10}
\bibitem{note_pn_limit}
For drawn here conclusions about features of dislocation structures
originating in lattice discreteness 
only the  character of the $F_{el} (d)$ and $F_{mis}(d)$ is important.
The oscillatory $F_{mis}(d)$ dependence  originates in lattice discreteness/periodicity.
Characteristic $F_{el} (d)$ dependence originates from the 
long wave length limit behavior of the
linear lattice response  Green function.
\bibitem{note_rig_derr}
This form can be  derived if (i) summation in Eq. \ref{eq:misf_ft} is
limited to the  s=1 oscillatory term
(a well justified approximation, since
this series is very rapidly converging ($J_{2}/J_{1} \ll 1 $));
(ii)  a dislocation density distribution function
represented as a sum of the partial dislocation densities,
$\rho(x) = \rho^{1}(x+d) + \rho^{2}(x-d)$; and (iii)
parity properties -   positive for $\rho(x) $ and  negative for  the
$\partial  \Phi ( u (x))/ \partial u$ functions are taken into account.
\bibitem{note_amp}
The estimate (based on the saddle point method)
allows to find  that
$A(d,\omega) \sim \exp (- \frac {2 \pi \omega }{h}) b_{k}H_{k}$, where $H_{k}$ 
are the main values of  the
$\gamma$-surface
curvature tensor at the unstable stacking fault point (H).
We also find that $\omega$ is the inverse proportional to the  $H$, thus
$A(d,\omega$) is
strongly dependent on  the geometrical ($H$)  and energetic
($\gamma_{isf}$) characteristics
of the $\gamma$-surface.
Details of the analysis of the relation between dislocation structure
and $\gamma$-surface geometry/energetics will be published elsewhere.
\bibitem{note_schoeck_func}
In the limit $d>>\omega$, $A(d,\omega) \approx A$ and derived in this paper
expression Eq.\ref{tot_nform} can be transformed to the one which similar
to  suggested by Schoeck \cite{schoeck_oss94}.
\end{thebibliography}

\begin{figure}[t]  
\caption{
Dependence of the generalized elastic and misfit  forces (in $J/m^{2}$) on
partials separation  $d$
(in lattice constant units) for
Ir
and CuAu
calculated within the PN model (left panel (a), (c)) and
for simple model step function shaped ordinary dislocation
  (right panels (b), (d)).
The misfit forces corresponding to the consistent discrete representation
of the misfit energy in the PN model are presented by solid lines and those
calculated within
the "continuum" approximation
 by dotted-dashed lines.
}
\label{fig:forces}
\end{figure}

\begin{figure}[t]
\caption{
Dislocation energy (in $J/m$) as (a) a function of the partials separation
$d$,
(b) a  corresponding  dependence of
$d$ on $l$ (for more than one stable core configuration,
solutions which are close in energy are numbered ``1''
and ``2'') and (c) a  corresponding dependence of
the total energy on the  position of the ordinary dislocation
center given by $l$ in units of a repeat  distance in the direction
normal to the dislocation line ($h$),
calculated for Ir (left panels) and  CuAu (right panels);
the ones determined within the "continuum" approximation (see text) are
presented by the dotted-dashed lines.
 }
\label{fig:compl}
\end{figure}

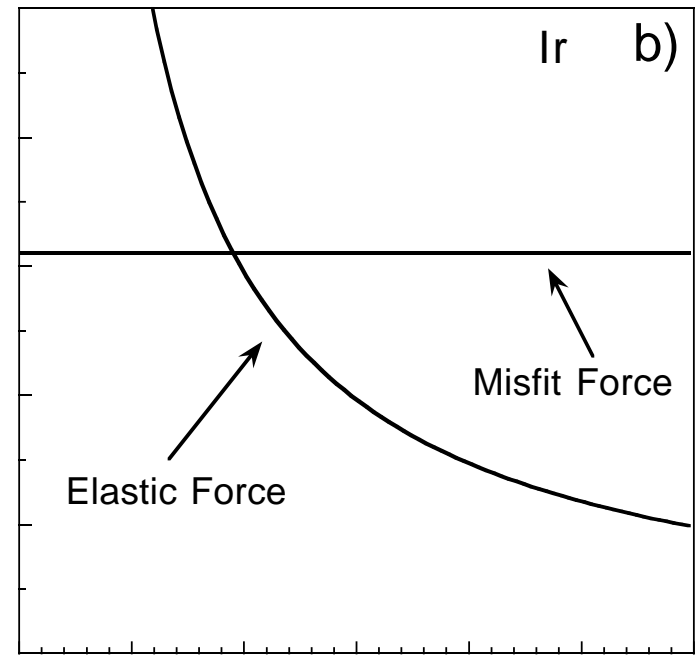
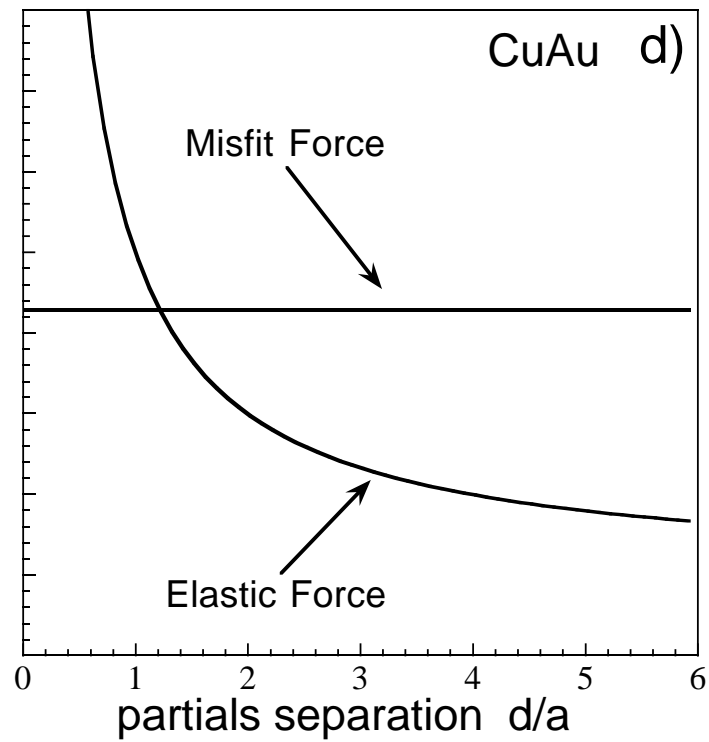
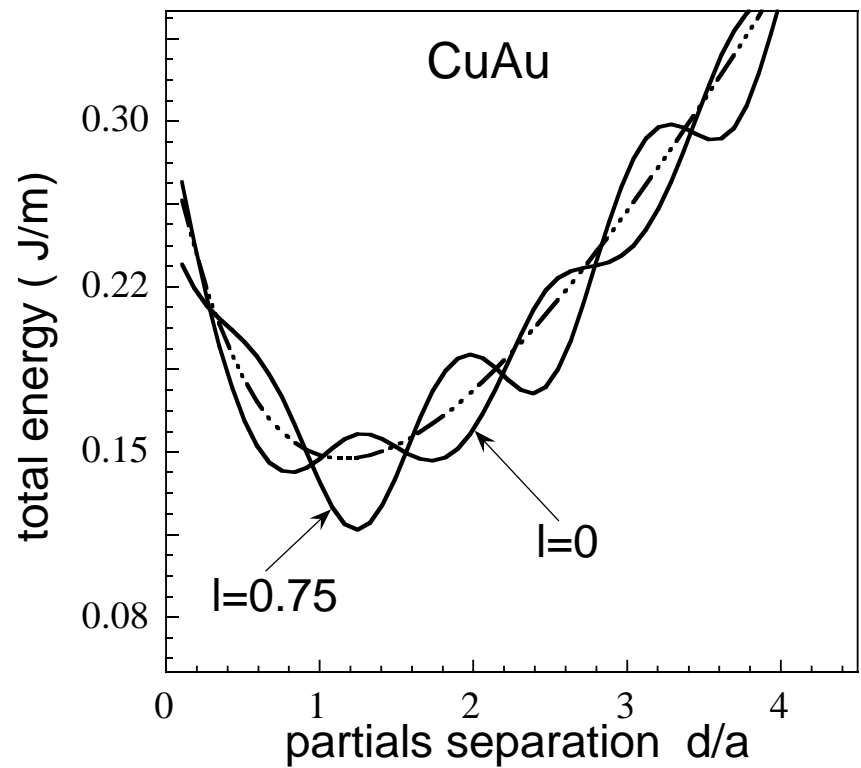
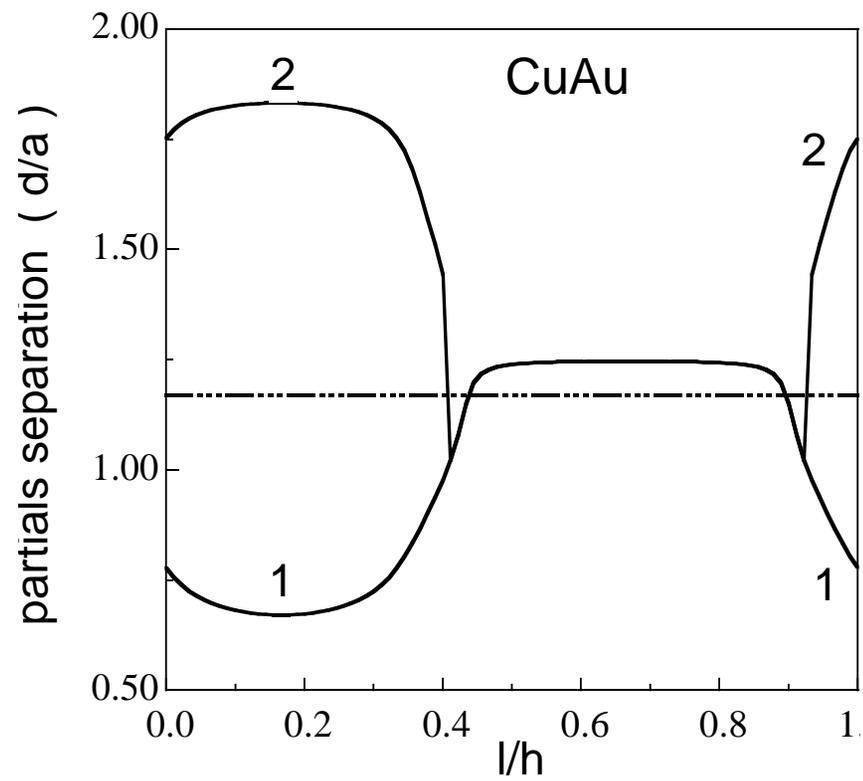
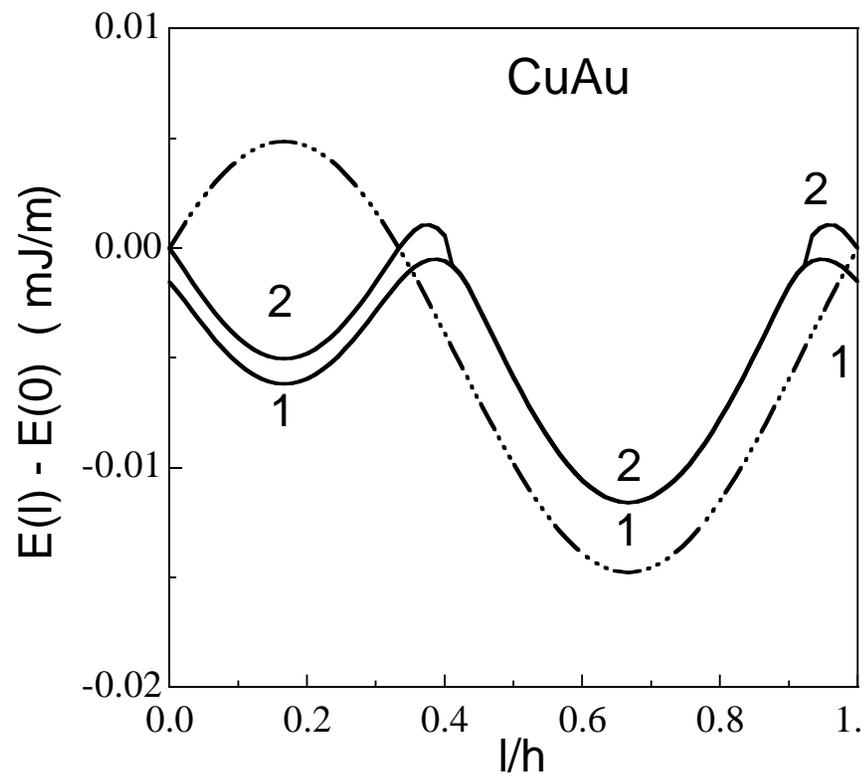
\end{document}